\documentclass{jfm}
\usepackage{graphicx}
\usepackage{epstopdf, epsfig}
\usepackage{float}
\usepackage{subfigure}
\usepackage{amsmath,amssymb}
\usepackage{natbib}
\usepackage{xcolor}
\usepackage{dcolumn}%
\usepackage{bm}%
\newcommand{\lra}[1]{\langle #1 \rangle }

\def\be{\begin{equation}}
\def\ee{\end{equation}}

\shorttitle{Stochastic modelling of acceleration}
\shortauthor{S. Chibbaro, A. Innocenti, N Mordant, N Stelzenmuller}
\title{Lagrangian stochastic modeling of acceleration in turbulent wall-bounded flows}
\author{Alessio Innocenti\aff{1}, 
  Nicolas Mordant\aff{2}
\and Nick~Stelzenmuller\aff{2},
Sergio~Chibbaro\aff{1}}

\affiliation{\aff{1} Sorbonne Universit\'e, CNRS, UMR 7190, Institut Jean Le Rond d'Alembert, F-75005 Paris, France
\aff{2} Laboratoire des Ecoulements G\'eophysiques et Industriels, Universit\'e Grenoble Alpes, CNRS, Grenoble-INP,  F-38000 Grenoble, France}

\begin{document}

\maketitle

\begin{abstract}
The Lagrangian approach is natural to study issues of turbulent dispersion and mixing. We propose in this work a general Lagrangian stochastic model including velocity and acceleration as dynamical variables for inhomogeneous turbulent flows.
%
The model takes the form of a diffusion process and
the coefficients of the model are determined via Kolmogorov theory and the requirement of consistency with the velocity-based models.
It is shown that the present model generalises both the acceleration-based models for homogeneous flows and the generalised Langevin models for the velocity.
The resulting closed model is applied to a channel flow at high Reynolds number  and compared to experiments as well as direct numerical simulations. 
A hybrid approach coupling the stochastic model with a Reynolds-Averaged-Navier-Stokes (RANS) is used to obtain a self-consistent model, as commonly used in probability density function methods.
Results highlight that most of the acceleration features are well represented, notably the anisotropy and the strong intermittency.
These results are valuable, since the model allows to improve the 
modelling of boundary layers yet remaining relatively simple.
It sheds also some light on the statistical mechanisms  at play in the near-wall region.
\end{abstract}

\section{Introduction}
Lagrangian stochastic models are widely used to describe complex turbulent flows~\citep{Pope_2000,fox2003computational,chibbaro2014stochastic}, and are of particular relevance for turbulent dispersion~\citep{wilson1996review}, reactive flows~\citep{Pope_1985,fox2003computational} and inertial particles~\citep{Minier_2001,Peirano_2006,minier2016statistical}.
They are also appealing when Lagrangian properties are under investigation~\citep{yeung1989lagrangian,mordant2002long,meneveau2011lagrangian,watteaux2019time}.
The stochastic models produce collections of synthetic trajectories that reproduce the statistical and, in some weak sense, the dynamical properties of particles advected by the flow.
In this framework, stochastic models focus generally on the dynamics of the one-point and one-time probability density function (pdf) of the state-vector of the system~\citep{Pope_2000}. 

The choice of the appropriate state-vector, that is to consider the  relevant variables for the model, is key to have an approximate but still accurate description of the turbulent flow considered. In particular, it is essential to retain enough variables 
to have a state-vector which can be described as Markovian~\citep{onsager1953fluctuations,gardiner1985handbook,marconi2008fluctuation}.
Since these models are mainly conceived to tackle general non-homogeneous flows, the state vector is usually limited to the position and velocity of the fluid particles $({\bf x}, {\bf U})$, and these observables are modelled as a diffusion process, which is justified by Kolmogorov theory~\citep{monin2013statistical,Pope_1994}.
The validity of such models may become questionable 
in some circumstances. 
In velocity-based models, the auto-correlation of the velocity is non-differentiable at the origin, reflecting the fact that the velocity process is not differentiable.
Moreover, the criterion used to include only fluid particle location and velocity in the state-vector and thus to neglect the behaviour at very high time-frequency  is to have separation of scales
$\tau_{\eta} \ll T_{L}$, where $\tau_{\eta}$ is the smallest time-scale, namely the Kolmogorov time-scale $\tau_{\eta}\equiv(\frac{\nu}{\lra{\epsilon}})^{1/2}$, and $T_L$ is the integral scale, typical of large-scale motion.
This criterion is no longer met at low Reynolds numbers,
 with the Reynolds number defined as $Re=\frac{UL}{\nu}$ where $U,L$ are typical velocity and length of large scales and $\nu$ is the kinematic viscosity. Mostly important it is not valid everywhere in a turbulent boundary layer.
In this case, there is no more the separation of scale justifying the Markovian description of the process $({\bf x}, {\bf U})$. 
Therefore, more variables should be added and in particular
the fluid particle acceleration. 

Some proposals have been made to address these issues in isotropic turbulence~\citep{krasnoff1971langevin,Saw_91}.
Furthermore, first experiments accessing Lagrangian fluid acceleration in isotropic flows~\citep{voth1998lagrangian,voth2002measurement,mordant2004experimental} have motivated many following attempts with the purpose of fitting the experimental data, but without a sound link with turbulence theory~\citep{gotoh2004turbulence}, except in few exceptions~\citep{lamorgese2007conditionally}.
An important step forward has been taken by \cite{Pope_2002}, who has 
proposed to use a general diffusion stochastic model for the fluid particle velocity and acceleration, and has considered in detail the case of homogeneous anisotropic turbulence.
In fact the purpose of that work was not to propose a specific model but 
rather to show that such a diffusion process may reproduce quite well the DNS data, if coefficients are correctly prescribed.
In particular, DNS was used to inform the model in that work.
Yet, no attempt has been made to develop a consistent model for non-stationary non-homogenous flows, although this is the more realistic and relevant situation for applications. 

The aim of this work is precisely to propose a first model including  the acceleration of fluid particles for the
general case of non-stationary and inhomogeneous turbulence, and it follows  recent accurate measurements and DNS of the acceleration in turbulent channel-flow at high Re-number~\citep{Nick}.
The model is developed in the general framework of a diffusion process for the fluid particle velocity and acceleration, as suggested by \cite{Pope_2002}. 
The coefficients are however specified not using data but on the basis of Kolmogorov theory and by the general analysis of the behaviour of the statistical moments (Reynolds stress) which can be extracted from the model.
In the simplest case of isotropic flows, the model reverts to the  model proposed earlier~\citep{krasnoff1971langevin,Saw_91}.
In non-homogeneous flows, when the acceleration can be considered as a fast process, that is when the observation scale is much larger than the characteristic one of the acceleration, which is related to the Kolmogorov one, the model is consistent instead with the standard Langevin models for the fluid velocity~\citep{Pope_2000}

The resulting model is therefore general and it can be used in all realistic flows of practical interest. 
We also regard the present model as an intermediate step in the development of a corresponding model for inertial particles.

\section{Theoretical Model}
As in the earlier works concerning isotropic turbulence \citep{krasnoff1971langevin,Saw_91}, we develop a model for the variables $({\bf x},{\bf U},{\bf a})$, which are respectively the position, the velocity and an additional random acceleration process. 
 For general non-homogeneous flows, we propose the following model:
\begin{eqnarray}
dx_{i}&=&U_{i}\, dt  
\label{model_fluidep} \\
dU_{i}&=& - \frac{1}{\rho}\frac{\partial \lra{P}}{\partial x_{i}}\, dt
+D_{ij}(U_{j}-\lra{U_{j}})\, dt
+ a_{i}\, dt, 
\label{model_fluidev} \\ 
da_{i} &=& -\beta a_{i} dt 
+ \sqrt{B}dW_{i},
\label{model_fluidea}
\end{eqnarray}
where $\langle P \rangle$ is the average pressure and $\lra{U_{i}}$ the average velocity.
As already remarked~\citep{Saw_91}, this form is equivalent to that including explicitly the full fluid acceleration $A_i = dU_i / dt$ (see appendix \ref{app:acc}) as an independent variable and is preferred here to highlight the differences with respect to stochastic models used for the reduced state vector including only the position and velocity of the fluid parcels $({\bf x},{\bf U})$.
It is important to underline here that several terms depend  on space $\mathbf x$, namely  $\langle P \rangle~,~D_{ij}~,~\lra{U_{i}}$; also the coefficients  $\beta~,~B$ are function of $\mathbf x$, as it will be clarified shortly.\\
To fix the unknown coefficients, the first requirement we ask to the model is to be consistent with the state-of-the-art Langevin model for non-homogeneous flows,
which has the following expression~\citep{Pope_2000}
\begin{eqnarray}
dX_i &=& U_i \,dt \nonumber \\ 
dU_i &=& -\frac{1}{\rho}\frac{\partial \lra{ P }}
{\partial x_i}\, dt +
G_{ij}\left( U_j- \lra{ U_j }\right)\, dt + \sqrt{C_0\lra{ \epsilon }}dW_i~,
\label{eq:modelv} 
\end{eqnarray}
with
$
G_{ij}=-\frac{1}{T_L}\delta_{ij}+G^a_{ij}\;\;$
where $G^a_{ij}$ is a traceless matrix,
$\text{with}\;\; T_{L}= { (\frac{1}{2}+\frac{3}{4}C_{0})}^{-1}
\frac{k}{\lra{\epsilon}}$,
where ${k}(\mathbf x)$ is the turbulent kinetic energy and $\lra{\epsilon}(\mathbf x)$ is the average dissipation-rate.
This condition fixes immediately $D_{ij}=G_{ij}$. 
The matrix $G^a_{ij}$ may be used to improve the quality of the Reynolds-stress model corresponding to the stochastic model. Since in this work we use a refined RANS model (to obtain average quantities) together with adding the effect of the acceleration, we want to assess the effect of this inclusion and therefore we take
 $G^a_{ij}=0$ for the sake of simplicity. 
 As it will be clear later, even with this approximation the model is satisfactory.  
Furthermore, looking at the two models,
we can see that our new model replaces the white-noise in the velocity Eq. (\ref{eq:modelv}), which is delta-correlated, with a coloured noise, that is a process $ a_i $ in Eq. (\ref{model_fluidev}) with a finite correlation time.
In particular, we have chosen to model $ a_i $ as a
Ornstein-Uhlenbeck  with correlation time given by $\beta^{-1}$, as expressed by eq. (\ref{model_fluidea}).
Therefore, the form given by (\ref{eq:modelv}) should be obtained in the limit of infinite $ \beta$, as explained later.

We use dimensional arguments \emph{\`a la Kolmogorov}  
to give an estimate of the time-scales.
Considering
$
\delta U_{\tau}=\vert \mathbf{U}(t+\tau)-\mathbf{U}(t)\vert,~
D^{L}(\tau)=\lra{ (\delta U_{\tau})^2}=v_{\eta}^2 \zeta(x)$, where $v_{\eta}\equiv(\nu \lra{\epsilon})^{1/4}$,
$x=\frac{\tau}{\tau_{\eta}}$, with $\tau_{\eta}\equiv\left(\frac{\nu}{\lra{\epsilon}}\right)^{1/2}$, and
$\zeta(x)$ is an universal function.
In the inertial range that gives $
D^{L}(\tau) \sim C_0\lra{\epsilon} \, \tau,
$ where $C_0$ is a constant. 
This implies for the velocity autocorrelation function that
$
R_L(\tau)\equiv\frac{\lra{U(t)U(t+\tau) }}{\overline{U^2}} = 
1- \frac{D^L(\tau)}{2\overline{U^2}} \sim 1-\frac{C_0}{2}\frac{\tau}{T}
$, with $T$ the time-scale of large scales, such that $R_L(\tau)\sim1$ in the inertial range far from boundaries 
when $\tau \ll T$~\citep{monin2013statistical,Pope_1994}.
We can generalise this result for the correlation of velocity derivatives 
$
B_{n}^{L}(\tau) = \lra{\frac{d^nu}{dt^n}(t)\frac{d^nu}{dt^n}(t+\tau)}
$
, where the same hypotheses yield 
$
B_{n}^{L}(\tau) = \left(\frac{v_{\eta}}{\tau_{\eta}^n}\right)^2 \alpha_n(x)
= \nu^{1/2-n}\lra{\epsilon}^{n+1/2}\alpha_n(x)~.$ 
Where 
$\alpha_n(x)$ should be  universal functions too. 
In particular, using the definition of $D^L$, we can find $B_{n}^{L}(\tau)=\frac{(-1)^{(n-1)}}{2} \frac{d^{2n} D^L}{d\tau^{2n}}$, that means $\alpha_n(x)=\frac{(-1)^{(n-1)}}{2} \frac{d^{2n}\zeta(x)}{dx^{2n}}$.
If $\tau$ is in the inertial range,
we 
demand that
the function $B_{n}^{L}$ should be independent of $\nu$.
We then obtain
$
B_{n}^{L}(\tau)\approx\lra{\epsilon}\tau^{1-2n}$, i.e., $\alpha_n(x) 
\sim x^{(1-2n)}
\label{k41}
$. 
Hence, for $n=1$ we find the acceleration correlation behaviour
$
R_A(\tau)=\frac{\lra{ A(t+\tau)A(t) }}{\lra{ A^2 }}
\sim \frac{\tau_{\eta}}{\tau} 
$, 
and for $n=2$ that of the derivative of acceleration
$
R_{\dot{A}}(\tau)\sim \left(\frac{\tau_{\eta}}{\tau}\right)^3 
$. 
These formulas show that the turbulent acceleration and
 higher derivatives are correlated on time comparable to $\tau_{\eta}$.
 Furthermore, the same formulas give for the mean square of the turbulent acceleration $\overline{A^2}=K\nu^{-1/2}\lra{\epsilon}^{3/2}$ where $K$ is a universal constant.
 This shows that acceleration, as well as its derivatives, depends on the fluid viscosity
$\nu$. 
Since for high-Re numbers the viscosity affects only 
the very small scales of turbulent motion, 
in locally isotropic turbulence these variables are 
 determined largely by scales $l \le \eta$.
On the basis of these estimates, the fluid particle
acceleration timescale $\beta^{-1}$ is taken proportional to the local Kolmogorov timescale, assuming
 a constant of proportionality of one we get
$
\beta^{-1}=\tau_{\eta}$.
Our rationale shows that the higher the derivative the faster the process loses correlation, so that higher order variables can be considered random noise. Nevertheless, for $\tau \lesssim \tau_{\eta}$ this reasoning ceases to be true and 
higher-order models may have some justification.

Once fixed the time-scale of the process ${\bf a}$, the consistency with the velocity-model (\ref{eq:modelv})
allows us to fix the diffusion term in (\ref{model_fluidea}).
From the equations~(\ref{model_fluidep})-(\ref{model_fluidea}), we can derive the corresponding
 Reynolds-stress equations   
\begin{equation}
\begin{split}
\frac{\partial \lra{ u_iu_j }}{\partial t}
+ \lra{ U_k }\frac{\partial \lra{ u_iu_j }}{\partial x_k}
+\frac{\partial \lra{ u_iu_ju_k }}{\partial x_k} =& 
-\lra{ u_iu_k }\frac{\partial \lra{ U_j }}
{\partial x_k} - \lra{ u_ju_k }\frac{\partial \lra{ U_i}}
{\partial x_k} \\
& + \lra{ u_ia_j } + \lra{ u_ja_i } - \frac{2}{T_L} \langle u_i u_j \rangle.
\end{split}
\end{equation}
where the correlations $\lra{ u_ia_j }$ are solutions of transport equations, which reflect the non-zero memory effects due to the coloured noise in the velocity.
Specifically, the transport equations for the $\langle u_i a_j \rangle$ correlation are
\begin{equation}
\frac{\partial \lra{ u_ia_j }}{\partial t}
+ \lra{ U_k }\frac{\partial \lra{ u_ia_j }}{\partial x_k}
+\frac{\partial \lra{ u_ia_ju_k }}{\partial x_k} = - \frac{1}{T_L}\lra{ a_ju_i }
+ \lra{ a_ia_j } 
-\lra{ a_ju_k } \frac{\partial \lra{ U_i}}
{\partial x_k} 
-\frac{\lra{ u_i a_j }}{\tau_{\eta}}~;
\label{eq:traspua}
\end{equation}
and for the variance of the process $a$
\begin{equation}
\frac{\partial \lra{ a_ia_j }}{\partial t}
+ \lra{ U_k }\frac{\partial \lra{ a_ia_j }}{\partial x_k}
+\frac{\partial \lra{ a_ia_ju_k }}{\partial x_k} = -2\frac{\lra{ a_i a_j }}{\tau_{\eta}}
+ B\delta_{ij}~.
\label{eq:traspaa}
\end{equation}
The finite value of timescale $\tau_{\eta}$ is responsable of the memory effect,
thus, in the limit of $\tau_{\eta} \rightarrow 0$, 
the same source term in Reynolds-stress equations as given by the model (\ref{eq:modelv}) should be found.
Considering the limit in the homogeneous case, we have 
$ 
\lra{a_ia_j} \rightarrow  \frac{B \tau_{\eta}}{2} \delta_{ij}\Rightarrow
\lra{u_ia_j} = \tau_{\eta}( \lra{a_ia_j}-\frac{\lra{u_ia_j}}{T_L})
,
$ 
and thus, $ \lra{u_ia_j} \rightarrow \frac{B}{2} \tau_{\eta} ( \frac{1}{\tau_{\eta}} + \frac{1}{T_L})^{-1}$;
therefore, in order to have the correct limit, it is necessary to impose
\begin{equation}
B=\frac{C_0 \lra{\epsilon}}{\tau_\eta}\left(\frac{1}{\tau_\eta}+\frac{{1}}{ T_L}\right)\approx\frac{C_0 \lra{\epsilon}}{\tau_\eta^2}
\label{eq:B}
\end{equation}
In this way, we have fixed all the parameters and the model is complete, except for the numerical choice of the free parameter $C_0$, which in this work is chosen to be $C_0=0.35$ to be consistent with the Rotta constant used in the RANS model~\citep{Pope_2000}. In any case, the results do not change qualitatively in the range $C_0\in[0.2,1.5]$.
Taking the trace of Reynolds-stress equations, 
the equation for turbulent kinetic energy is retrieved.
As discussed in~\cite{pope2014determination},
the dissipation rate is implicitly defined by the relation 
$
\sum_i \lra{ u_iA_i }\approx -\lra{\epsilon}, \;
\label{eps_def}
$
(where $A_i = d u_i / dt$, see appendix \ref{app:acc}).

\section{Numerical method}
We study the turbulent flow in a channel between two parallel walls separated by a distance $2h$ using the same Reynolds number ($Re_{\tau} =\frac{u_{\tau} h}{\nu}\approx 1440$) chosen in a recent campaign of experiments and DNS~\citep{Nick}, where $u_{\tau}$ is the friction velocity associated to the shear stress $\tau_w$ at the wall and $\nu$ the kinematic viscosity. In the following, the superscript $+$ indicates quantities expressed in wall units by $u_{\tau}$ and $\nu$. By convention, $x$ is the stream-wise direction, $y$ the wall-normal and $z$ the span-wise.

We must solve the stochastic differential equations (\ref{model_fluidep})-(\ref{model_fluidea}) that contain several average fields. 
To cope with this issue, we use here a hybrid approach~\citep{Pope_1985,Pope_2000} computing the needed average quantities ($\lra{\mathbf U}, \lra{P}, \lra{\epsilon}, k$) solving the Reynolds averaged Navier-Stokes (RANS) equations. 
In particular,  we have implemented the near-wall model based on the elliptic relaxation \citep{durbin1991near},  using the same parameters and boundary conditions suggested in the original proposal~\citep{durbin1993reynolds}.
These Reynolds-stress equations together with the equation for $\lra{\epsilon}$ are solved through a standard finite-difference method. 

Concerning the stochastic equations (\ref{model_fluidep})-(\ref{model_fluidea}),
they are of the form $d{\bf X}={\bf A} \,dt+ {\bf B}{\bf X} \,dt + {\bf D} \,d{\bf W}(t)$ and are stiff, as we have $\displaystyle \lim_{y\rightarrow 0} det [{\bf B}]=-\infty$. 
Indeed, $T_L^{-1}\sim - \frac{\epsilon}{k}$ and near to the wall this means $T_L^{-1}\sim\frac{-1}{y^2}$, which grows without bound. Moreover near the wall 
$U_i-\lra{U_i} \sim y$, for the no-slip boundary condition. Thus, the drift coefficient $[{\bf B}]{\bf U}$ scales with $1/y$ and remains unbounded for $y\rightarrow 0$.
Furthermore, the time-scale $\beta$ may be very small and instabilities may arise also far from boundaries if the time-step is bigger (typically in the inertial range) and the numerical scheme is not stable.
To address these issues, we have developed a numerical scheme unconditionally stable for the set of equations (\ref{model_fluidep})-(\ref{model_fluidea})~\citep{Dreeben_1998,Peirano_2006}.
The numerical scheme is presented in the appendix \ref{app:a}.

\section{Results}
\begin{figure}
\centering
\includegraphics[scale=0.25]{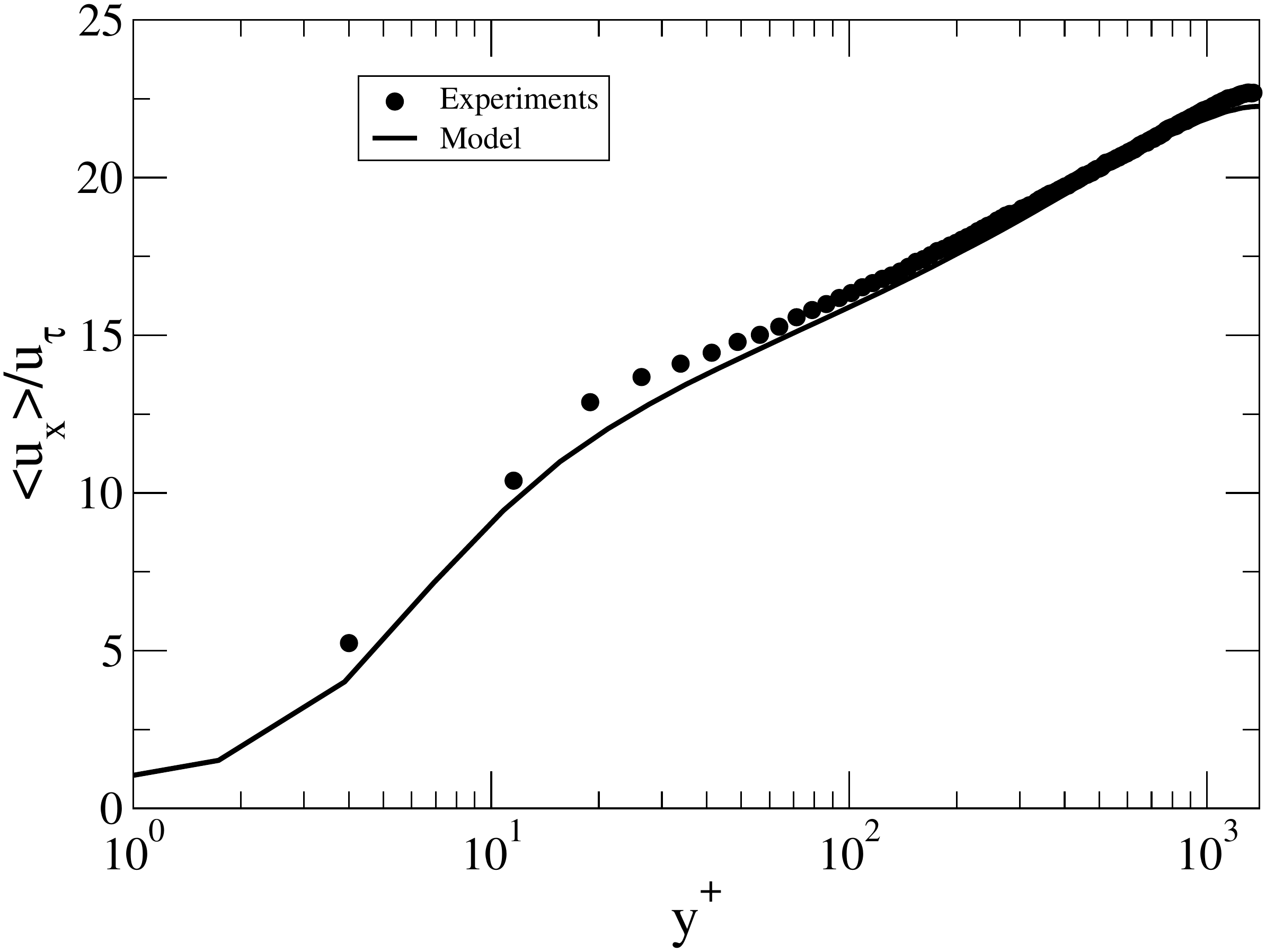}
\includegraphics[scale=0.25]{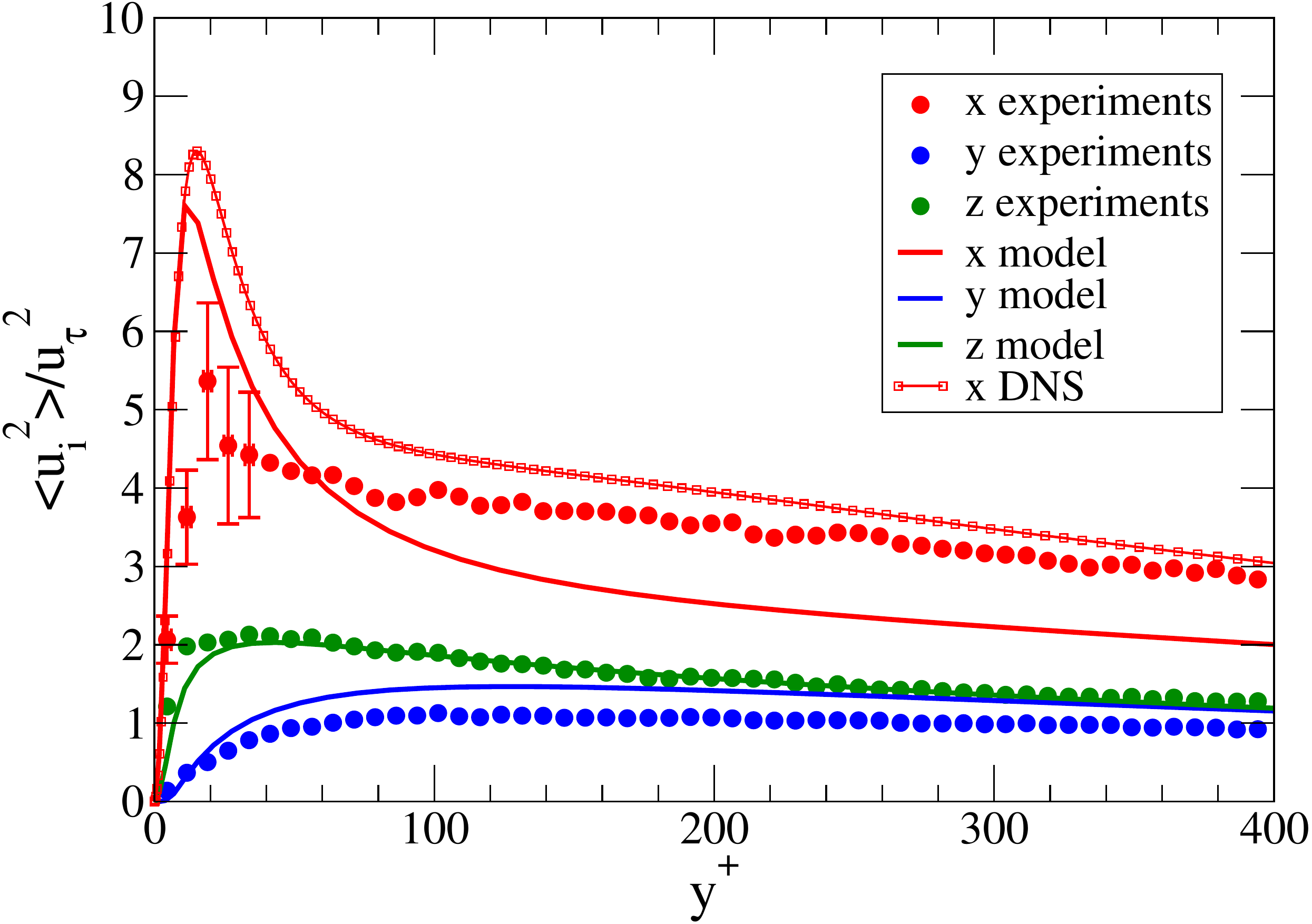}
\caption{Mean (a) and variance velocity profiles (b). Comparison between experiments (points) and the present model (lines). 
All quantities are normalized in wall units.
}
\label{fig1}
\end{figure}
In Fig. \ref{fig1}, mean and variance velocity profiles from experiments and model simulations are compared. 
Given that we use a hybrid approach, the Reynolds stress are computed in the model by the RANS method.
The agreement of average velocity is very good. The Durbin model gives also a satisfactory representation of the Reynolds stress flow. 
It is known that  RANS models tend to underestimate the stream-wise component in the log-layer region in very high-Re flows~\citep{Pope_2000}.
Nonetheless, we are not interested here to develop further RANS models, and the chosen model is satisfactory with regard to the present purpose.
\begin{figure}
\centering
\includegraphics[scale=0.25]{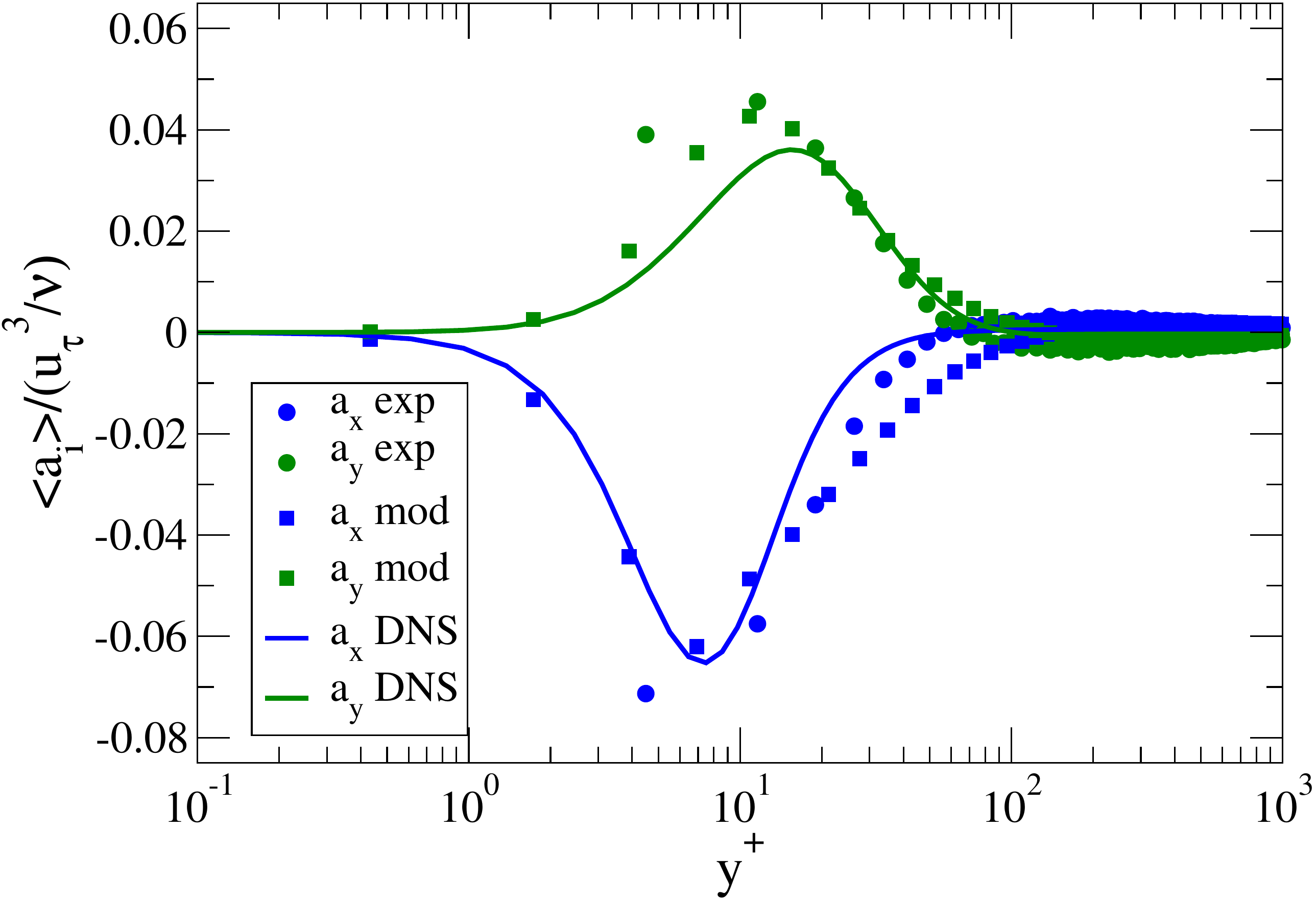}\includegraphics[scale=0.25]{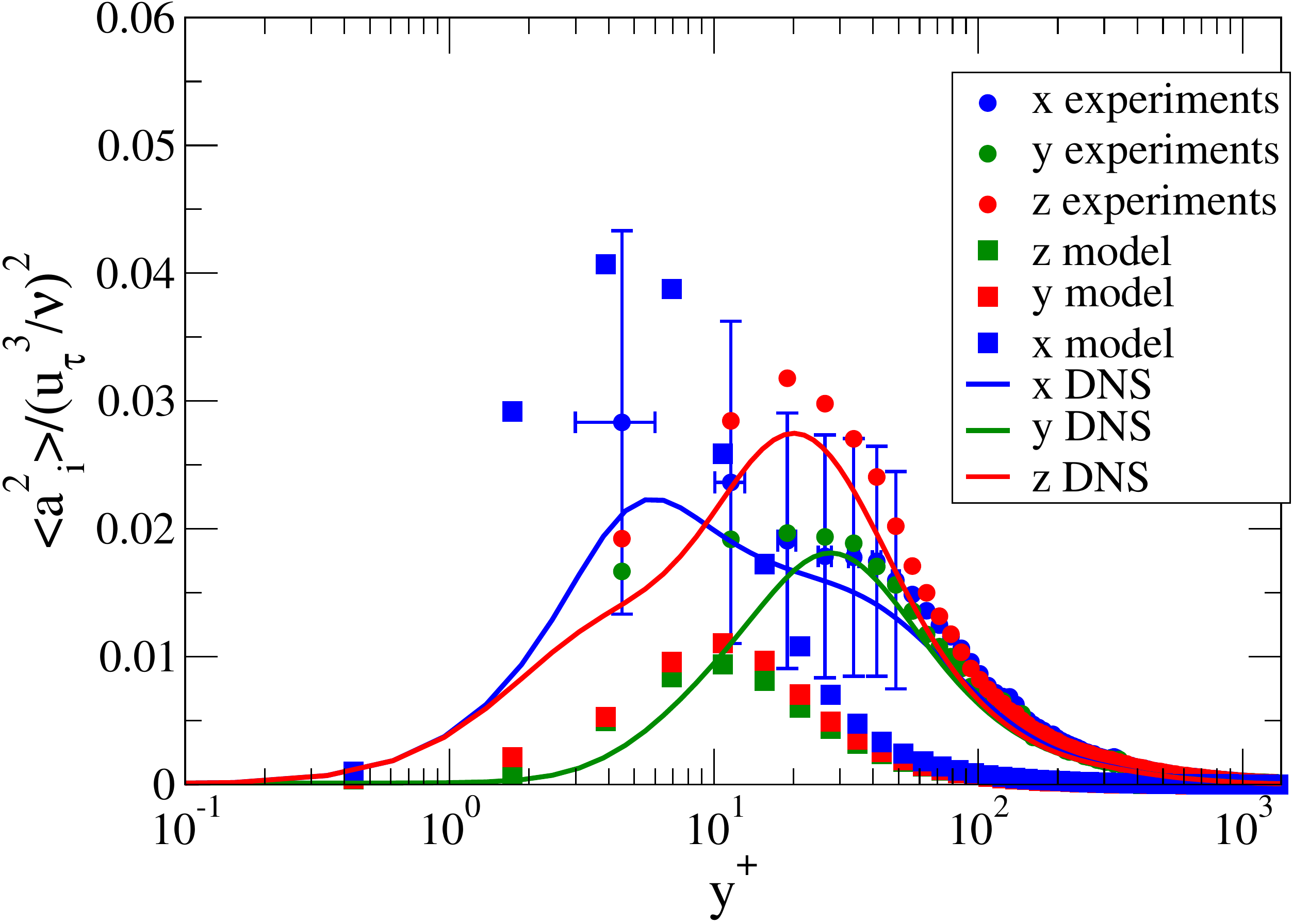}
\caption{Mean and variance acceleration profiles. Comparison between experiments (points), stochastic model simulations (squares) and DNS (lines).}
\label{fig2}
\end{figure}
Figure \ref{fig2} shows the mean and variance acceleration profiles obtained by the experiments, DNS and the model. 
The average acceleration is very well predicted by the stochastic model in both directions.
Notably the model gives correctly
the negative peak of mean stream-wise acceleration at $y^+ \approx 7$, which is a viscous effect. That shows that the acceleration model together with a RANS model including the boundary layer is able to describe this effect, despite the absence of ad-hoc low-Re terms in the stochastic model.
Profiles of acceleration variance (Fig. \ref{fig2}) reveal a qualitative overall agreement, although larger discrepancies are found for the model concerning the variance accelerations. 
In particular, a slight overestimation of the stream-wise variance is present, even though the value found is within the experimental error bars. More pronounced, the model displays  the peak of the spanwise and cross-stream components around the same position, whereas experiments and DNS show some variability.
Nevertheless, at their respective peaks, the standard deviation of acceleration is larger than the magnitude of the mean acceleration
for all sets, indicating that the present stochastic model is already able to reproduce the main features of the acceleration fluctuations that govern the dynamics near the wall. 

\begin{figure}
\centering
\includegraphics[scale=0.225]{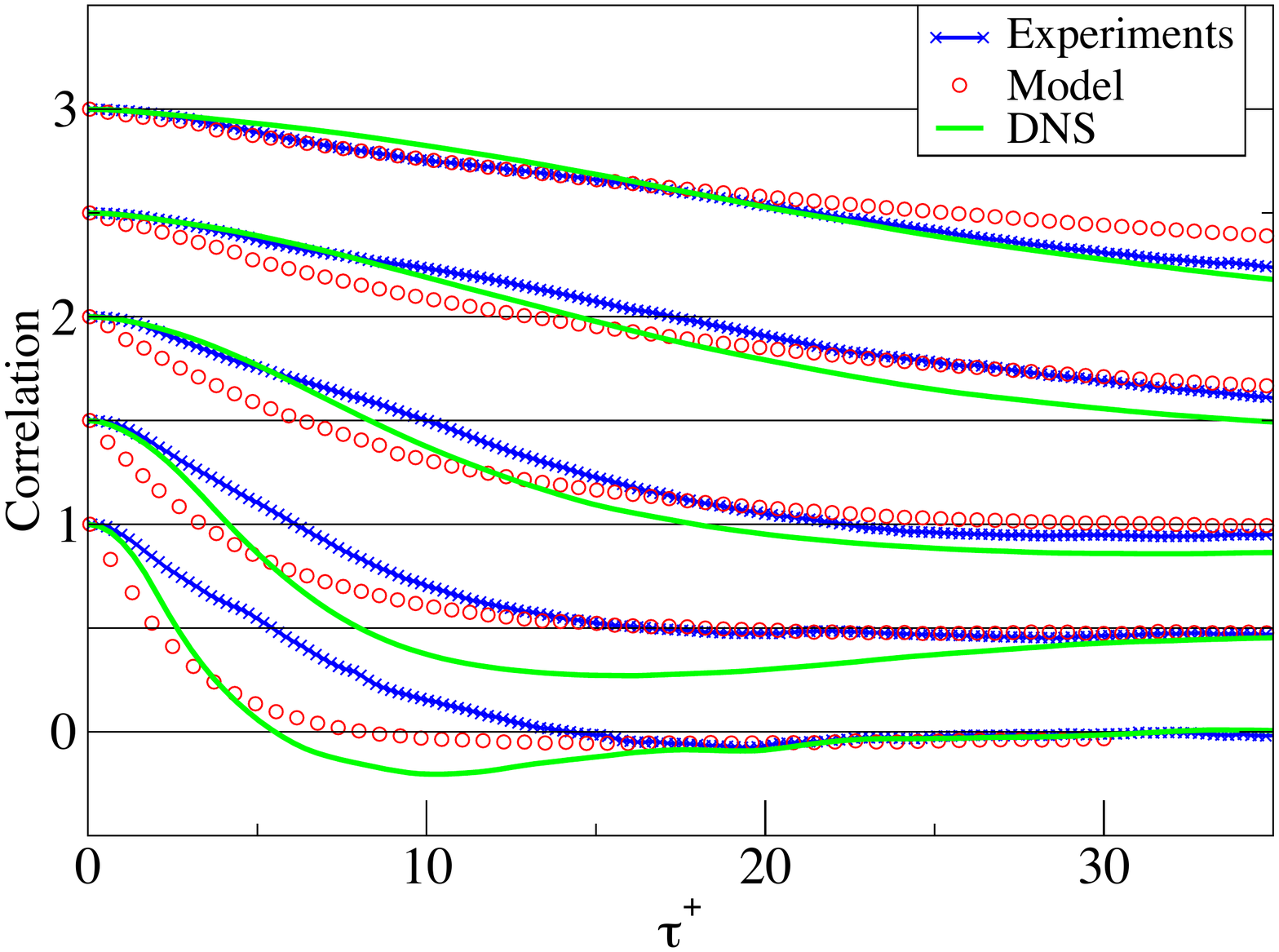}
\includegraphics[scale=0.225]{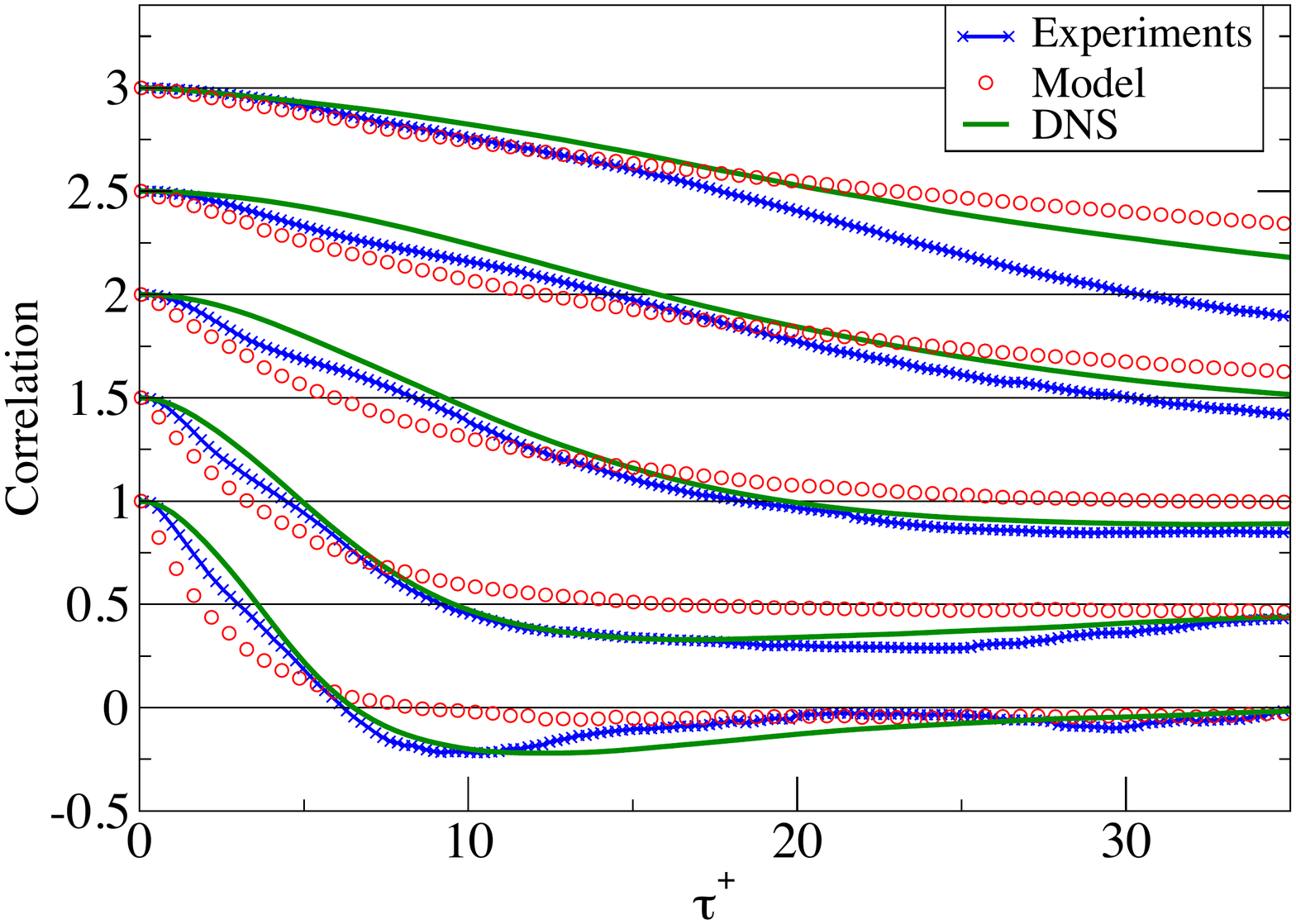}
\caption{Lagrangian auto-correlations of wall-normal ($\rho_{yy}$, left) and spanwise ($\rho_{zz}$, right) particle acceleration. 
Correlations are computed as:
$\rho_{ij}(\tau,y_0) = \frac{\lra{a_i^\prime(t_0,y_0) a_j^\prime(t_0+\tau,y_0)}}
{\lra{a_i^{\prime 2}(t_0,y_0)}^{1/2} \lra{a_j^{\prime 2}(t_0+\tau,y_0)}^{1/2}}$.
Experiments crossed blue lines, model red lines. 
Curves are shifted vertically by increments of 0.5 for clarity. From bottom to top, the curves correspond to particles located initially at $y_0^+ = 20, 60, 200, 600$ and $1000$.
Horizontal grid lines show the zero-correlation level for each $y_0^+$.
The DNS data curve is shown for one case, for comparison.}
\label{fig3}
\end{figure}
Figures \ref{fig3}a-b show $y$ and $z$ components of the acceleration correlation tensor $\rho_{ij}$ calculated at different initial wall
distances $y_0^+$. 
The agreement is satisfactory, showing that the stochastic model fairly reproduce both the inhomogeneity and the anisotropy of the flow, since all components are different and the decorrelation-time changes with the distance.
Some small differences for small time-displacements, notably in the near-wall region, can be traced back to the presence of a white noise in the acceleration process which overlooks short-memory effects.

\begin{figure}
\centering
\includegraphics[scale=0.35]{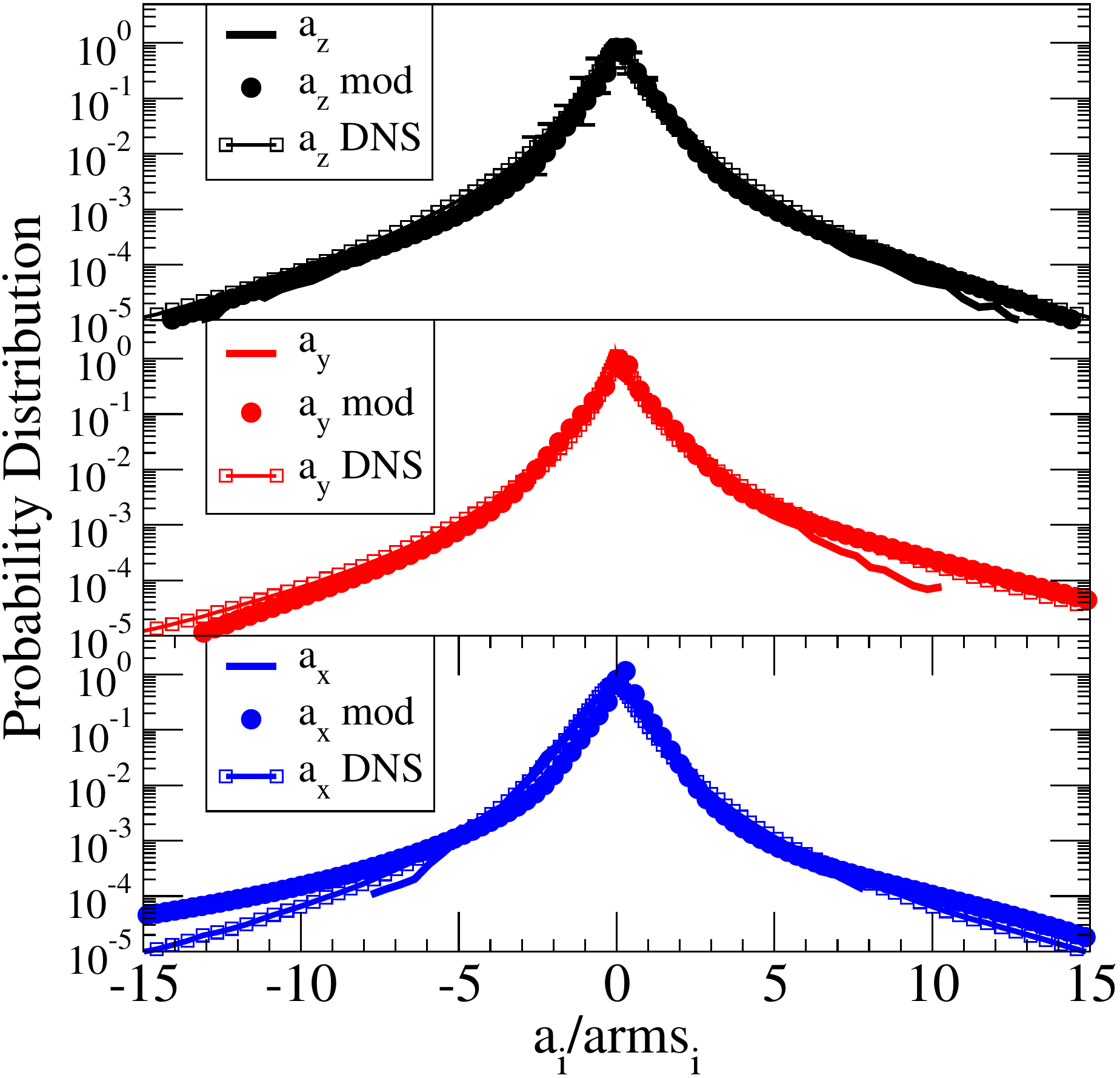}\\
\caption{PDF of streamwise, wall-normal and spanwise particle acceleration. Experiments - solid lines. Model - symbols. 
The PDFs are normalized by the root-mean-square value of acceleration. The PDF represent an average over the whole channel.}
\label{fig4}
\end{figure}
Figure \ref{fig4} shows the probability distribution function (PDF) of the three acceleration components. All curves present very long tails corresponding to extremely high acceleration events usually associated to intermittency (\cite{mordant2002long}).
Once again, the model reproduces well the experimental and DNS results.
In particular, the model captures well the skewness of the acceleration, displaying a positive skewness for the cross-stream component and a negative one for the stream-wise one. 
The tails of the extreme events ($P<10^{-4}$) appear slightly over-predicted by the model for the stream-wise component. However statistical error in this range is significative.

\section{Discussion and Conclusions}
In this work, we have developed a novel stochastic model including fluid particle acceleration for general non-homogeneous turbulent flows.
This model generalises both previous propositions for acceleration in isotropic flows and the velocity-based models for non-homogeneous flows, and is compared against experiments and DNS in a channel-flow.
To deal with the fluid averages in the coefficients, 
 we have used a hybrid RANS/PDF approach, typically used in realistic computations,
 where average velocity and Reynolds stress are computed in a Eulerian framework.
In particular, we have chosen the elliptic-relaxation Reynolds-stress model to get a fair agreement between the RANS fields and the experimental ones.
As for the proposal of the stochastic model, it is derived using Kolmogorov theory and the moment equations for the covariance of the velocity and acceleration.
The use of this hybrid formulation with the inclusion of the acceleration should bring some of the effects related to the viscous terms, and present results are also useful to disentangle which effects are obtained in this way and which remains to be introduced with more complex closures.
The overall behaviour of the model is remarkable showing that it captures most of the features revealed by experiments, and notably is capable to model correctly the anisotropy of the flow.
The average acceleration is in agreement with experiments and DNS. 
The Lagrangian autocorrelation given by the model reproduces correctly the time-scale and the non-homogeneous effects.

Interestingly, also the pdfs are fairly well captured, displaying skewness, anisotropy and far-from-gaussian tails. In homogeneous turbulence, the very wide tails of the acceleration PDF are associated to intermittency (\cite{Mordant:2001}). In order to reproduce these features in stochastic models of homogeneous turbulence, a multiplicative component is added in the noise term to take into account the fluctuations of the dissipation rate $\epsilon$ that are often assumed to follow log-normal statistics and to have a rather long correlation time close to $T_L$(\cite{Pope_1990,mordant2002long,Reynolds:2003,Zamansky:2010ik}). Our model does not have such a multiplicative term in eq. (\ref{model_fluidea}). However, the term $B$ incorporates the dissipation rate $\lra{\epsilon}$ which varies in space along the particle trajectory due to the inhomogeneity of the flow. It is thus a multiplicative term only related to inhomogeneity and not on stochastic fluctuations of $\epsilon$. It appears here that this contribution is enough to reproduce the wide tail PDFs of the acceleration components.
This confirms on one hand that in wall flows most of the extreme events are related to non-homogeneity~\citep{lee2004intermittent},
and on the other hand shows that the present model is able to capture the main statistical features of the near wall structures 
which are responsible of the intermittency.

In order to enhance further the performances, improvements would be:
(a) to add intermittency effects in the turbulent dissipation~~\citep{Pope_2000} and/or in the acceleration~\citep{lamorgese2007conditionally};
(b) to improve the velocity model with a non-null matrix $G_{ij}^a$ in eq. (\ref{model_fluidev}),  which would allow to retrieve full consistency~\citep{chibbaro2011note,minier2014guidelines} and further improve the Reynolds-stress. 
In conclusion, we believe that the present form of the stochastic model is adequate to model wall flows in realistic case and should be the starting point to develop an acceleration-based model for inertial particles.
\appendix
\section{Numerical scheme}
\label{app:a}
Writing equations (\ref{model_fluidep})-(\ref{model_fluidea})
as $d{\bf X}={\bf A} \,dt+ {\bf B}{\bf X} \,dt + {\bf D} \,d{\bf W}(t)$, we take the matrix coefficients ${\bf A},{\bf B},{\bf D}$ frozen during a time-step $\Delta t$ to obtain analytical solutions using the integrating factor $\displaystyle e^{-{\bf B}t}$.
Therefore, the scheme reads as:
\begin{align}
& \quad x_{i}^{n+1} = x_{i}^n + A_1\,U_{i}^n + B_1\,f_{i}^n
  + C_1\,[T_i^n C_i^n] + \Omega _i^n,
   \\
& \quad U_{i}^{n+1} = U_{i}^n\, \exp(-\Delta t/T_i^n)
              + [T_i^n C_i^n] [1-\exp(-\Delta t/T_i^n)]
              + \gamma _i^n, \\
& \quad a_{i}^{n+1} = a_{i}^n\, \exp(-\Delta t/\tau^n)
              + D_1\,U_{i}^n + [T_i^n C_i^n](E_1-D_1)
              + \Gamma _i^n.   \\   \notag 
& \quad A_1 = \tau^n\,[1-\exp(-\Delta t/\tau^n)],
\quad B_1 = T_i^n/(T_i^n-\tau^n) \,[T_i^n(1-\exp(-\Delta t/T_i^n)-A_1], \notag \\
&   \quad C_1 = \Delta t - A_1 - B_1, \quad D_1 = T_i^n/(T_i^n-\tau^n) [\exp(-\Delta t/T_i^n)-\exp(-\Delta
  t/\tau^n)],\notag\\  
& \quad E_1 = 1 - \exp(-\Delta t/\tau^n).\notag \\ 
& \quad \gamma _i(t) = \Check{B}_i\exp(-t/T_i)
  \int _{t_0}^{t} \exp(s/T_i)\,dW_i(s),   \notag \\
& \quad \Gamma _i(t) = \frac{1}{\tau}\exp(-t/\tau)
  \int _{t_0}^{t}\exp(s/\tau)\,\gamma _i(s)\,ds,   \notag \\ 
& \quad \Omega _i(t) = \int _{t_0}^{t}\Gamma
_i(s)\,ds.  \notag
\end{align}

\section{Expression of the second-order model} \label{app:acc}
The present model may be written as a second order-process
considering the entire acceleration as the main variable~\citep{Saw_91}:
\begin{eqnarray}
dx_{i} &=& U_{i} \, dt \\
dU_{i} &=& A_{i}\, dt \\
dA_{i} &=& -\left( \frac{1}{2}+\frac{3}{4}C_{0}\right) \frac{\lra{\epsilon}}{k}\left(A_{i}-\lra{A_{i}}-\frac{U_i}{\tau_\eta}\right)\, dt  \notag \\
&+& \frac{1}{\tau_\eta} \left( A_i+\frac{1}{\rho} \frac{\partial \lra{P}}{\partial x_{i}} \right)\, dt
-\frac{d}{dt}\frac{1}{\rho}\frac{\partial \lra{P}}{\partial x_{i}}  
+\sqrt{\frac{C_0 \lra{\epsilon}}{\tau_\eta}(\frac{1}{\tau_\eta}+\frac{{1}}{ T_L})}dW_{i}.
\end{eqnarray}

\bibliographystyle{jfm}
\bibliography{biblio}

\end{document}